\documentclass[10pt,conference]{IEEEtran}
\IEEEoverridecommandlockouts
\usepackage{cite}
\usepackage{amsmath,amssymb,amsfonts}
\usepackage{algorithmic,latexsym}
\usepackage{graphicx}
\usepackage{textcomp,url,stackrel, float}
\usepackage{xcolor,bm}
\usepackage{epsfig}
\usepackage{epstopdf}
\def\BibTeX{{\rm B\kern-.05em{\sc i\kern-.025em b}\kern-.08em
    T\kern-.1667em\lower.7ex\hbox{E}\kern-.125emX}}
\newcommand{\EE}{\mathsf{E}}
\usepackage[normalem]{ulem}
\begin{document}

\title{Blind Adaptive Equalization in Additive Impulsive Noise Using the Logarithmic Product Fractional-Moment (LP-FM) Criterion}

\author{\IEEEauthorblockN{Shafayat Abrar}
\IEEEauthorblockA{Dhanani School of Science and Engineering,\\ Habib University, 
Karachi 75290, Pakistan \\
\textbf{Email:} \color{blue}{\textbf{shafayat.abrar@sse.habib.edu.pk}}}
}

\maketitle

\begin{abstract}
The blind mitigation of inter-symbol interference in additive white impulsive noise modeled by a symmetric $\alpha$-stable (S$\alpha$S) distribution is investigated. A novel logarithmic product fractional-moment statistics (LP-FMS) criterion is proposed by combining complementary fractional-moment statistics with logarithmic normalization in a constrained optimization framework. Based on this criterion, a normalized blind equalization algorithm for symmetric alpha-stable noise (NBEA-SAS) is derived using stochastic gradient ascent with recursive fractional-moment estimation and Bussgang-consistent constrained adaptation. Simulation results for $64$-APSK signaling over fractionally spaced multipath microwave channels show that the proposed algorithm converges faster than FLOS-CMA, RAW-CMA, and NBEA-GG while achieving a comparable steady-state residual intersymbol interference floor under both moderately and highly impulsive S$\alpha$S noise conditions. The results demonstrate that the proposed LP-FMS criterion provides a robust framework for blind adaptive equalization in impulsive noise environments.
\end{abstract}


\section{Introduction}
Impulsive noise, characterized by a heavy-tailed distribution, is a major impairment in digital communication systems. Its principal sources include electromagnetic induction, electrostatic coupling, lightning, and other forms of radio-frequency interference \cite{Shao1993}. Communication systems designed under the Gaussian noise assumption suffer severe performance degradation in impulsive environments. Blind equalization mitigates intersymbol interference (ISI) introduced by multipath propagation without requiring knowledge of either the channel response or the transmitted sequence. Instead, it exploits the statistical properties of the received signal and additive noise \cite{HaykinBook1994,ding2001blind}. Blind equalizers optimize a cost function so that the statistics of the equalized output match those of the transmitted signal. Consequently, effective operation in impulsive environments requires cost functions that explicitly account for both the transmitted signal and additive noise.

The first comprehensive investigation of blind equalization in additive impulsive noise was reported by Mathis and Douglas \cite{mathis2003bussgang}. They showed that most Bussgang-type blind equalizers are inadmissible and prone to divergence unless the equalizer norm is constrained. Since the required constraint depends on the unknown optimal equalizer, the approach is impractical.

To address this limitation, Rupi \textit{et al.} \cite{rupi2004constant} proposed the \textsl{fractional lower-order statistics constant modulus} (\textrm{FLOS-CM}) cost function for symmetric alpha-stable (S$\alpha$S) noise:
\begin{equation}\label{EqRupiCost}
\begin{aligned}
&\EE\big||y_k|^{p-1}y_k-R\,y_k\big|^q
=\EE\left[\big|y_k\big|^q\cdot\big||y_k|^{p-1}-R\big|^q\right],\\
& p<\alpha,\;\; q<\alpha.
\end{aligned}
\end{equation}
Here, $y_k$ is the equalizer output and $R$ is the \textit{dispersion radius}. Since an S$\alpha$S random variable possesses finite moments only for orders below $\alpha$, the parameters $(p,q)$ are selected accordingly. However, simulations were limited to mildly impulsive noise (\(1.85\leqslant\alpha<2\)).

Li \textit{et al.} \cite{li2018robust} later proposed the \textsl{robust adaptive weighted constant modulus} (\textrm{RAW-CM}) algorithm based on \textit{zero-order statistics} (ZOS):
\begin{equation}\label{RAWcost}
\EE\log\!\left(1+\big(|y_k|^p-R\big)^2\right).
\end{equation}
Its stochastic-gradient implementation employs a normalization factor that suppresses the influence of impulsive outliers. 
The ZOS framework, introduced by Gonzalez \textit{et al.} \cite{gonzalez2006zero}, relies on the logarithmic moment \(\EE\log|X|\), which remains finite for a broad class of heavy- and light-tailed distributions. Consequently, it applies to both finite- and infinite-variance models, including S$\alpha$S, generalized Cauchy, and Pareto distributions.

More recently, Abrar \textit{et al.} \cite{abrar2020adaptive} investigated blind equalization in generalized Gaussian noise. Using Wijsman's theorem from statistical invariance theory, they derived the cost:
\begin{equation}\label{EqRelaxedCost}
\bm{w}^\dagger=
\stackbin[\bm{w}]{}{\textrm{arg}}\,\textrm{max}\,
\EE|y_k|^p
\quad
\textrm{s.t.}
\ p\leq\alpha,\quad
|y_k|\leqslant R\,\,\,\forall\, k.
\end{equation}
The resulting normalized blind equalization algorithm (NBEA-GG) converged substantially faster than RAW-CM \cite{li2018robust} and FLOS-CM \cite{rupi2004constant,tang2009capture} over several multipath channels in generalized Gaussian noise.

Despite these advances, important limitations remain. FLOS-based algorithms lose robustness under highly impulsive conditions, whereas ZOS-based methods do not exploit the convergence benefits of product-type cost functions. Moreover, NBEA-GG is derived for generalized Gaussian noise and does not naturally extend to S$\alpha$S processes. These limitations motivate the proposed blind equalization algorithm.

\section{Proposed Method}

The objective of this work is to develop a blind adaptive equalization algorithm for additive white symmetric alpha-stable (S$\alpha$S) noise by combining the complementary strengths of FLOS-CMA, RAW-CMA, and NBEA-GG. Rather than optimizing a single fractional moment, the proposed formulation maximizes the logarithm of the product of two complementary fractional moments. This product criterion captures the evolution of the equalizer-output distribution more effectively during adaptation, while logarithmic normalization suppresses the influence of impulsive outliers, resulting in faster and more robust convergence.

We first introduce the logarithmic product fractional-moment (LP-FM) criterion and then derive the corresponding adaptive algorithm, referred to as the \emph{normalized blind equalization algorithm for symmetric alpha-stable noise} (NBEA-SAS). The LP-FM criterion possesses three key features:
\begin{enumerate}
\item \textit{Complementary fractional moments:} Combines complementary fractional moments to balance robustness against impulsive noise with sensitivity to the equalizer-output distribution.
\item \textit{Logarithmic normalization:} Compresses large outliers, reducing their influence on the adaptive update.
\item \textit{Constrained optimization:} Restricts the equalizer output modulus to an admissible region.
\end{enumerate}

The LP-FM criterion is formulated as
\begin{equation}\label{EqRelaxedCost2}
\begin{aligned}
\bm{w}^{\dagger}
&=
\stackbin[\bm{w}]{}{\textrm{arg}}
\,\textrm{max}\,
\log\!\left(
1+\EE|y_k|^p\,\EE|y_k|^q
\right)
\\
\textrm{s.t.}\quad
&
|y_k|\leqslant R,\quad \forall\,k,
\quad
p\leqslant\alpha,\quad
q\leqslant 2-p.
\end{aligned}
\end{equation}
Unlike FLOS-CM, the proposed criterion maximizes the logarithm of the product of complementary fractional moments. The logarithmic mapping attenuates the influence of impulsive outliers, while the product formulation preserves the rapid convergence associated with fractional-moment-based blind equalization.

To derive the adaptive algorithm, (\ref{EqRelaxedCost2}) is optimized using stochastic gradient ascent. Throughout the derivation we select \(p=\alpha\) and \(q=2-p\), and define
\[J:=
\log\!\left(
1+\EE|y_k|^p\,\EE|y_k|^{2-p}
\right).\]
Its gradient is
\begin{align*}
\frac{\partial J}{\partial\bm{w}_k^\ast}
=
\frac{1}
{1+\EE|y_k|^p\,\EE|y_k|^{2-p}}
\,
\frac{\partial
\left(
\EE|y_k|^p\,\EE|y_k|^{2-p}
\right)}
{\partial\bm{w}_k^\ast}.
\end{align*}

Applying the product rule gives
\begin{align*}
\frac{\partial
\left(
\EE|y_k|^p\,\EE|y_k|^{2-p}
\right)}
{\partial\bm{w}_k^\ast}
&=
\EE|y_k|^{2-p}
\,
\frac{\partial\EE|y_k|^p}
{\partial\bm{w}_k^\ast}
\\
&\quad+
\EE|y_k|^p
\,
\frac{\partial\EE|y_k|^{2-p}}
{\partial\bm{w}_k^\ast}.
\end{align*}

The ensemble expectations are approximated by exponentially weighted iterative estimates,
\begin{align}
\widehat{\EE}|y_k|^{2-p}
&\leftarrow
\widehat{\EE}|y_k|^{2-p}
+
c
\left(
|y_k|^{2-p}
-
\widehat{\EE}|y_k|^{2-p}
\right),
\\
\widehat{\EE}|y_k|^p
&\leftarrow
\widehat{\EE}|y_k|^p
+
c
\left(
|y_k|^p
-
\widehat{\EE}|y_k|^p
\right),
\end{align}
where \(c\ll1\) is the forgetting factor. Hence,
\begin{align*}
\frac{\partial
\left(
\EE|y_k|^p\,\EE|y_k|^{2-p}
\right)}
{\partial\bm{w}_k^\ast}
&\approx
\widehat{\EE}|y_k|^{2-p}
\frac{\partial|y_k|^p}{\partial\bm{w}_k^\ast}
+
\widehat{\EE}|y_k|^p
\frac{\partial|y_k|^{2-p}}{\partial\bm{w}_k^\ast}.
\end{align*}

Using Wirtinger calculus \cite{Adali2010},
\begin{align*}
\frac{\partial|y_k|^p}
{\partial\bm{w}_k^\ast}
=
\frac{p}{2}
\frac{y_k^\ast\bm{x}_k}
{|y_k|^{2-p}},
\quad\text{and}\quad
\frac{\partial|y_k|^{2-p}}
{\partial\bm{w}_k^\ast}
=
\frac{2-p}{2}
\frac{y_k^\ast\bm{x}_k}
{|y_k|^p}.
\end{align*}

Substitution yields
\begin{align}\label{eq_gradient_of_J}
\frac{\partial J}
{\partial\bm{w}_k^\ast}
=
\frac{
p
\dfrac{\widehat{\EE}|y_k|^{2-p}}
{|y_k|^{2-p}}
+
(2-p)
\dfrac{\widehat{\EE}|y_k|^p}
{|y_k|^p}
}
{
2
\left(
1+
\widehat{\EE}|y_k|^p
\widehat{\EE}|y_k|^{2-p}
\right)
}
\,y_k^\ast\bm{x}_k.
\end{align}

When \( |y_k|<R \), gradient ascent is performed,
\[
\bm{w}_{k+1}
=
\bm{w}_k
+
\frac{\mu}{\|\bm{x}_k\|^2}
\frac{\partial J}{\partial\bm{w}_k^\ast},
\]
where \(\mu>0\) is the step size.

If \( |y_k|>R \), the adaptation direction is reversed to force the output back into the admissible region,
\[
\bm{w}_{k+1}
=
\bm{w}_k
-
\zeta
\frac{\mu}{\|\bm{x}_k\|^2}
\frac{\partial J}{\partial\bm{w}_k^\ast}.
\]

The two modes are combined as
\begin{subequations}\label{EqAlg}
\begin{align}
\bm{w}_{k+1}
&=
\bm{w}_k
+
\frac{\mu}{\|\bm{x}_k\|^2}
\,g(y_k)\,
\frac{\partial J}
{\partial\bm{w}_k^\ast},
\\
g(y_k)
&=
\left\{
\begin{array}{rl}
1,
&
|y_k|<R,
\\
-\zeta,
&
\textrm{otherwise},
\end{array}
\right.
\end{align}
\end{subequations}
where \(g(y_k)\) switches automatically between ascent and descent.
The penalty factor \(\zeta\) is selected using the Bussgang theorem \cite[Eq.~(2.29)--(2.30)]{HaykinBook1994}, giving
\begin{equation}\label{EqValueofg}
\zeta
=
\frac{
4M
\EE|a|^p
\EE|a|^{2-p}
}{
M_L
\left[
(2-p)
R^{2-p}
\EE|a|^p
+
pR^p
\EE|a|^{2-p}
\right]
}
-1,
\end{equation}
whose derivation is provided in Appendix.

\medskip
\noindent
\textit{Remark~1:}
The normalization terms
\({\widehat{\EE}|y_k|^p}/{|y_k|^p}\) and
\({\widehat{\EE}|y_k|^{2-p}}/{|y_k|^{2-p}}\),
reduce the adaptation gain when impulsive samples exceed their running averages and increase it otherwise, thereby suppressing outliers while accelerating convergence. Together with the logarithmic normalization factor,
\(1+
\widehat{\EE}|y_k|^p
\widehat{\EE}|y_k|^{2-p}\),
the switching function \(g(y_k)\), and the term \(y_k^\ast\bm{x}_k\), they maximize the cost while ensuring
\(|y_k|\leqslant R,\, \forall\,k\) (that is, enforced in a statistical/steady-state sense).

\medskip
\noindent
\textit{Remark~2:}
Although
\(\widehat{\EE}|y_k|^p\)
and
\(\widehat{\EE}|y_k|^{2-p}\)
may be initialized to arbitrary non-negative values, simulations show that zero initialization consistently provides faster convergence and superior transient performance.

\section{Simulation Results}

Simulations are conducted to evaluate the proposed equalizer, hereafter referred to as \textit{normalized blind equalization algorithm for symmetric alpha-stable noise} (NBEA-SAS). Its performance is compared with the robust adaptive weighted constant modulus algorithm (RAW-CMA) \cite{li2018robust}, the fractional lower-order statistics constant modulus algorithm (FLOS-CMA) \cite{tang2009capture}, and the normalized blind equalization algorithm for generalized Gaussian noise (NBEA-GG) \cite{abrar2020adaptive}.

The simulations employ $64$-APSK signaling over fractionally spaced channels corrupted by additive white symmetric alpha-stable [S$\alpha$S$(\alpha,\gamma)$] noise. The channel impulse responses are sampled at twice the symbol rate, resulting in $T/2$-spaced processing with separate even- and odd-phase FIR equalizers. Both equalizers are initialized with a single central tap set to unity and all remaining coefficients to zero. The propagation channels are selected from the SPIB microwave channel database \cite{SPIBLink}; among the fifteen available channels, channel-$1$ and channel-$2$ are considered.
The S$\alpha$S noise samples are generated using Weron's method \cite{Weron1995}. The in-phase and quadrature noise components are assumed mutually uncorrelated, while the noises affecting the even and odd receiver branches are assumed statistically independent.

For NBEA-SAS, NBEA-GG, RAW-CMA, and FLOS-CMA, the fractional-order parameter is set to $p=\alpha$. The characteristic exponent is assumed known \textit{a priori} or estimated during an initial idle period \cite{kateregga2017parameter}. For FLOS-CMA, $q=0.2$, and the dispersion radius $R$ for FLOS-CMA and RAW-CMA is determined from the Bussgang condition. Step sizes are listed in the figure legends. Equalizer lengths are $21$ and $41$ taps for channels-$1$ and $2$, respectively.
Convergence is evaluated using the intersymbol interference ratio (ISR) under two impulsive noise conditions,
\((\alpha,\gamma)=(1,10^{-2})\)
and
\((\alpha,\gamma)=(0.5,10^{-4})\),
representing moderately and highly impulsive environments, respectively.

\begin{figure}[htbp]\centering
\includegraphics[scale=0.32]{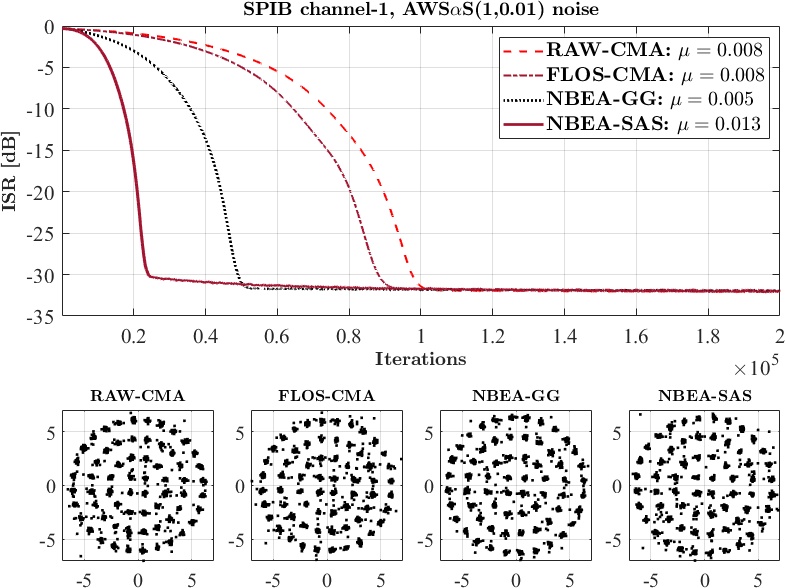}
\caption{SPIB channel-$1$ in S$\alpha$S\((1,1\times10^{-2})\) noise ($21$ taps).}\label{FigAllFSE-SASa}
\end{figure}

\begin{figure}[htbp]\centering
\includegraphics[scale=0.32]{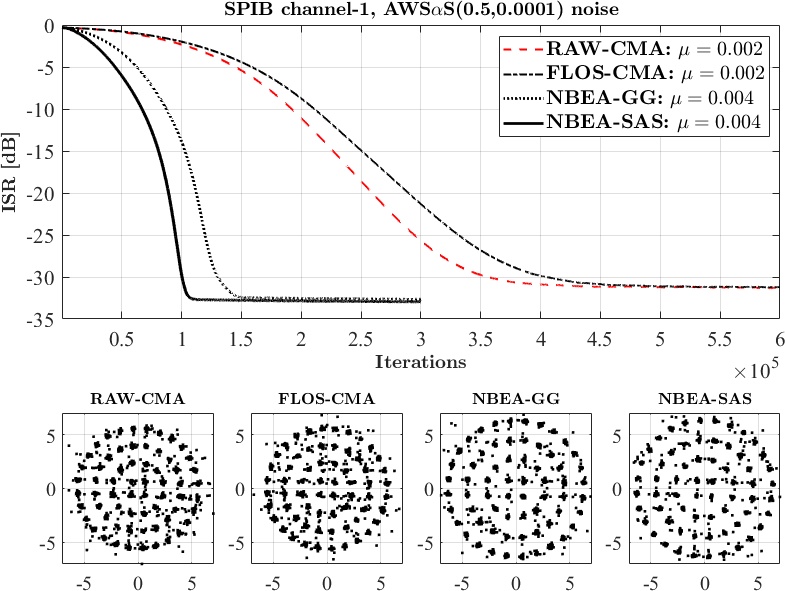}
\caption{SPIB channel-$1$ equalization in S$\alpha$S\((0.5,1\times10^{-4})\) noise ($21$ taps).}\label{FigAllFSE-SASb}
\end{figure}

\begin{figure}[htbp]\centering
\includegraphics[scale=0.32]{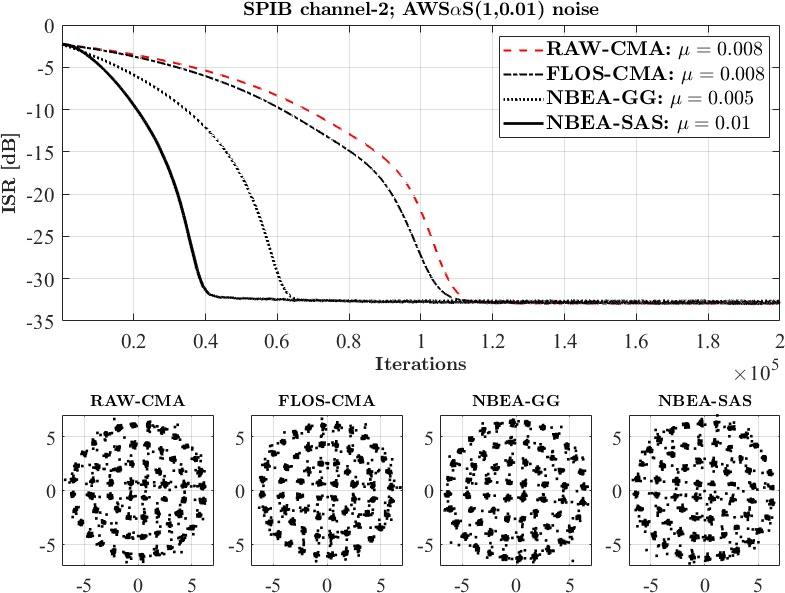}
\caption{SPIB channel-$2$ equalization in S$\alpha$S\((1,1\times10^{-2})\) noise ($41$ taps).}\label{FigAllFSE-SASc}
\end{figure}

To improve numerical stability under severe impulsive noise, two mechanisms are incorporated into all algorithms. Each quadrature component of the regressor vector $\bm{x}_k$ is clipped to $[-\tau,\tau]$ with $\tau=21$, and all coefficient updates are normalized by $\|\bm{x}_k\|^2$. These measures are applied uniformly to ensure a fair comparison.
Figures~\ref{FigAllFSE-SASa}--\ref{FigAllFSE-SASc} summarize the results. Each ISR curve is averaged over $100$ independent Monte Carlo trials using identical transmitted symbol sequences and identical S$\alpha$S noise realizations for all algorithms.
The proposed NBEA-SAS consistently exhibits the fastest convergence for both microwave channels under both impulsive noise conditions while achieving a steady-state ISR comparable to NBEA-GG, RAW-CMA, and FLOS-CMA. Its advantage is particularly evident in the highly impulsive case ($\alpha=0.5$), demonstrating that the combination of logarithmic normalization and product fractional-moment statistics provides both strong robustness against impulsive disturbances and rapid convergence.

\section{Conclusions}

This paper addressed blind channel equalization in additive white impulsive noise modeled by a symmetric alpha-stable (S$\alpha$S) distribution. A novel blind equalization criterion was developed by combining complementary fractional-moment statistics with logarithmic normalization in a constrained optimization framework. The resulting formulation exploits the convergence advantages of product-type statistical measures while retaining the robustness of logarithmic normalization. Based on this criterion, a normalized blind equalization algorithm for symmetric alpha-stable noise (NBEA-SAS) was derived using stochastic gradient ascent.

Simulation results for $64$-APSK signaling over fractionally spaced multipath microwave channels showed that NBEA-SAS converges faster than NBEA-GG, RAW-CMA, and FLOS-CMA while achieving a comparable steady-state residual intersymbol interference (ISI) floor. The improvement is particularly pronounced under highly impulsive S$\alpha$S noise, where conventional blind equalization algorithms are most susceptible to performance degradation.
Overall, the proposed LP-FMS criterion provides a robust framework for blind adaptive equalization in symmetric alpha-stable noise.

\section*{Appendix: Derivation of the Penalty Factor $\zeta$}

The penalty factor $\zeta$ is determined from the Bussgang equilibrium condition,
\(\EE\,\bm{w}_{k+1}=\EE\,\bm{w}_k\),
using the update
\[
\bm{w}_{k+1}
=
\bm{w}_k
+
\frac{\mu g(y_k)}
{\|\bm{x}_k\|^2}
\frac{
p\dfrac{\widehat{\EE}|y_k|^{2-p}}{|y_k|^{2-p}}
+
(2-p)\dfrac{\widehat{\EE}|y_k|^p}{|y_k|^p}
}
{
1+\widehat{\EE}|y_k|^p
\widehat{\EE}|y_k|^{2-p}
}
y_k^\ast\bm{x}_k.
\]

At convergence, the equalizer is assumed to satisfy the open-eye condition ($y_k=a_k$). The transmitted symbols are zero mean and mutually uncorrelated, additive noise is neglected, and the received signal is
\(
x_k=\sum_i h_i a_{k-i},
\)
where $h_0\neq0$. Since constant scaling does not affect the equilibrium condition, $\mu$ and $\|\bm{x}_k\|^2$ are omitted. Hence,
\[
0=
\EE\!\left[
\frac{
p\dfrac{\widehat{\EE}|y_k|^{2-p}}{|y_k|^{2-p}}
+
(2-p)\dfrac{\widehat{\EE}|y_k|^p}{|y_k|^p}
}
{
1+\widehat{\EE}|y_k|^p
\widehat{\EE}|y_k|^{2-p}
}
g(y_k)y_k^\ast x_k
\right]_{y_k=a_k}.
\]
Assuming the recursive moment estimates have converged to their ensemble values, define
\(S_1=p\,\EE|a_k|^{2-p}\),
and
\(S_2=(2-p)\,\EE|a_k|^p\).
Then
\[
0=
\EE\!\left[
\left(
S_1|a_k|^{p-2}
+
S_2|a_k|^{-p}
\right)
g(a_k)a_k^\ast x_k
\right].
\]
Substituting
\(
x_k
\)
and using symbol independence gives
\[
0=
h_0
\EE\!\left[
\left(
S_1|a_k|^p
+
S_2|a_k|^{2-p}
\right)
g(a_k)
\right].
\]
The bipolar switching function is
\[
g(a_k)=
\begin{cases}
1,
&
|a_k|<R,
\\
-\zeta,
&
|a_k|\ge R.
\end{cases}
\]
Substituting $g(a_k)$ yields
\begin{align}\label{eqIntermediateS1S2}
\nonumber
0
&=
\EE\!\left[
S_1|a_k|^p
+
S_2|a_k|^{2-p}
\right]_{|a_k|<R}
\\
&\quad
-
\zeta
\EE\!\left[
S_1|a_k|^p
+
S_2|a_k|^{2-p}
\right]_{|a_k|\ge R}.
\end{align}
At the boundary $|a_k|=R$, steady-state jitter is assumed equally likely to produce updates in either direction, giving
\begin{equation}\label{eqIntermediateS1S2A}
0
=
\EE[\cdots]_{|a_k|<R}
+
\frac12
\EE[\cdots]_{|a_k|=R}
-
\frac{\zeta}{2}
\EE[\cdots]_{|a_k|=R}.
\end{equation}
This approximation is consistent with small stochastic fluctuations around equilibrium.

For an $M$-symbol constellation with modulus levels
\[
|a_k|
\in
\{R_1,R_2,\ldots,R_L\},
\qquad
0<R_1<\cdots<R_L=R,
\]
let $M_j$ denote the number of points on the $j$th ring. Then
\(\EE|a_k|^p
=
\tfrac1M
\sum_{j=1}^{L}
M_jR_j^p.\)
Adding and subtracting the boundary contribution,
\[
\frac12
\frac{M_L}{M}
\left(
S_1R^p
+
S_2R^{2-p}
\right),
\]
reduces (\ref{eqIntermediateS1S2A}) to
\begin{align}\label{eqIntermediateS1S2B}
\nonumber
0
&=
S_1\EE|a_k|^p
+
S_2\EE|a_k|^{2-p}
-
\frac12
\frac{M_L}{M}
\left(
S_1R^p
+
S_2R^{2-p}
\right)
\\
&\quad
-
\frac{\zeta}{2}
\frac{M_L}{M}
\left(
S_1R^p
+
S_2R^{2-p}
\right).
\end{align}

Finally, substituting the definitions of $S_1$ and $S_2$ into (\ref{eqIntermediateS1S2B}) and solving for $\zeta$ yields (\ref{EqValueofg}).

\bibliographystyle{unsrt}
\bibliography{biblo}
\end{document}